\documentstyle[aps,prl]{revtex}
    \begin{document}
    \draft
    \title{
    Harmonic Vibrational Excitations in Disordered Solids and the "Boson Peak"
    }
    \author{Walter Schirmacher}
    \address{Physik--Department E13, Technische Universit\"at M\"unchen,
D--85747 Garching, Germany}
    \author{Gregor Diezemann}
    \address{Institut f\"ur physikalische Chemie, Universit\"at Mainz
Welderweg 15, D 55099 Mainz, Germany}
    \author{Carl Ganter}
    \address{Physik--Department E13, Technische Universit\"at M\"unchen,
D--85747 Garching, Germany}
    \maketitle
    \begin{abstract}
We consider a system of coupled classical harmonic oscillators with
spatially fluctuating nearest-neighbor force constants on a simple cubic
lattice.
The model is solved both by numerically diagonalizing the
Hamiltonian and by applying the single-bond coherent potential approximation.
The results for the density of states 
$g(\omega)$ are in excellent agreement with each other.
As the degree of disorder is increased the system becomes unstable due
to the presence of negative force constants. If the system is
near the borderline of stability
a low-frequency peak appears in the reduced
density of states
$g(\omega)/\omega^2$ as a precursor of the instability.
We argue that this peak is the analogon of the
"boson peak", observed in structural glasses.
By means of the level distance statistics we show that the peak is
not associated with localized states.
    \end{abstract}
A ubiquitous and rather intriguing feature in the physics of glasses
is the anomalous behavior of the low-frequency part of the vibration
spectrum and the corresponding thermal properties \cite{flm59,zp71,and81}.
While the origin of the linear low-tempature specific heat
is commonly attributed to the existence of double-well potentials or
two-level systems \cite{bk96}, there is still considerable debate about the
so-called "boson peak". This peak shows up in the density of states
(DOS) $g(\omega )$ as an excess contribution, compared to the usual
Debye behaviour ( $g(\omega )\propto \omega ^2$).
It is taken as being responsible for a number of features observed in
specific heat and thermal conductivity measurements, as well as
in Raman or neutron scattering data 
\cite{flm59,zp71,and81,bnd84,whl94,mwp97,dpr93}.
The boson peak seems also to persist at elevated temperatures. Here
its relation to the liquid-glass transition and the corresponding relaxation
dynamics
is a matter of discussion \cite{ret97,bbt97,sns97,gs96,fgm97}.
Due to the development of new experimental techniques allowing to
perform Brillouin scattering measurements in the THz range
\cite{bos95,mrs96,fcv96,bkm96},
as well as pertinent molecular dynamics simulations 
\cite{ls91,bl95,so96,cjj96,gg97}
the question concerning the nature of the modes in the
boson peak region has gained much additional interest recently.

Several models have been formulated to explain the physical origin
of the boson peak.
In the soft-potential model \cite{kki83,bgg92}
the existence of anharmonic localized
vibrations is postulated to be an intrinsic and essential property
of amorphous systems. The scattering of propagating phonons
by these modes is then taken as the origin for the excess DOS
in the boson peak region.

Another, almost orthogonal approach considers harmonic degrees of freedom
solely. Here the scattering of phonons is associated with
the disorder in the force constants, or, equivalently,
spatially fluctuating
elastic constants \cite{ell92}.
The first contribution in this direction has been
the phonon-fracton approach \cite{ao82,nyo94},
in which harmonic vibrational excitations on a
percolating lattice were considered.
However, a numerical simulation of this model \cite{nyo94}
showed that the crossover from propagating modes (phonons)
to localized fractal excitations (fractons) does not lead to
an excess DOS, although calculations using the
coherent potential approximation (CPA) \cite{doy84,pe88}
had predicted such an excess. 
Moreover, there is no experimental evidence for
spatially fractal behaviour of bulk glasses.

More realistic disordered coupled harmonic oscillator models
work in terms of a continuous distribution
of force constants.
Calculations of the vibration spectrum of such a model with a
specific force constant distribution have been performed
in effective medium
approximation (EMA, amorphous version of the CPA)
\cite{ws89,ws92,ws93,gan96}. In these calculations
an excess contribution in the DOS appears
in the frequency regime, where the phonon mean free
path becomes comparable to the wavelength.
However,
for this model the existence of the excess DOS has not yet been checked
numerically.

Another interesting contribution to the boson peak discussion \cite{gs96}
comes from the
mode-coupling theory \cite{got91}
of the liquid-glass transition. In the idealized
glass state a maximum in the density fluctuation spectrum at finite
frequencies is predicted which is associated with inhomogeneously
broadened harmonic oscillations. Experimental measurements exhibiting
a crossover from relaxational to vibrational dynamics \cite{whl94} 
have been described
successfully in terms of this theory \cite{fgm97}

Calculations for harmonic vibrational excitations in liquids, performed
numerically \cite{sk89,bl95}, as well in EMA \cite{wl92},
can hardly contribute to the present debate, because
the Hamiltonians considered are instable \cite{comment1}

In the present letter we investigate the properties
of a simple model for vibrational excitations of disordered systems
and show that excess low-frequency modes are present if the system
is almost unstable.
The model consists of a set of coupled
scalar harmonic oscillators placed on a three-dimensional
simple cubic lattice with lattice
constant $a\!=\!1$. The coupling among the oscillators
is modelled by nearest-neighbor force constants $K_{ij}$,
which are treated as independent (quenched) random variables,
chosen according to a distribution with density $P(K_{ij})$.
The corresponding Hamiltonian has off-diagonal elements
${\cal H}_{ij}=K_{ij}$ and diagonal elements
${\cal H}_{ii}=-\sum_{j\neq i}K_{ij}$.
For a stable system all eigenvalues
$\lambda_i=-\omega_i^2$ are negative, 
$\omega_i$ denoting the vibrational eigenfrequencies.
In all calculations we have chosen a truncated
Gaussian distribution with
\mbox{$P(K)=P_0\exp\{-(K-K_0)^2/2\sigma^2\}\theta(K-K_{\mbox{{\footnotesize min}}})$}.
Here $\theta (x)$ denotes the step function, $P_0$
is a normalization constant, $K_0 $ and
$\sigma $ denote the maximum value and the width, respectively.
In our calculations we set $K_0=1$.
The lower cut-off of the force constant distribution
is denoted by $K_{\mbox{{\footnotesize min}}}$.
This cut-off is introduced to allow for the study
of strongly disordered systems with a restricted
amount of negative force constants.

The model defined in this way has been analyzed, both, numerically
by diagonalizing the Hamiltonian matrix, as well as in
CPA.
The DOS obtained from both methods are in excellent
agreement with each other, except near the point
of instability.

For the numerical treatment we considered a cubic box of
size $L$ and imposed periodic boundary conditions.
The resulting $L^3\times L^3$ Hamiltonian matrix
was diagonalized using the NAG-LAPACK routine DSYEV.
The distribution of eigenvalues exhibits gaps
due to finite-size effects.
In order to eliminate these effects we calculated the
integrated density of levels
$F_L(\lambda)=\sum_i\theta(\lambda-\lambda_i)$
for a given size $L$, smoothed this function and averaged it
over different sizes, ranging from $L=10$ to $L=14$.
This procedure yielded a function $\bar F(\lambda)$,
the derivative of which gives the density of levels
$n(\lambda)= \mbox{d}\bar F(\lambda)/\mbox{d}\lambda$.
From this the vibrational DOS follows as
$g(\omega )= 2\omega n(-\omega^2)$.

For an approximate solution of our model
we have used the single-bond CPA \cite{ol81,web81,sum81}.
This theory is formulated
in terms of a frequency-dependent force constant ("self energy")
$\Gamma (\omega)$, which can be visualized as the inverse of an acoustic
dielectric function. The frequency-dependent complex sound velocity
$v(\omega)$ is given by $\Gamma (\omega)=v(\omega)^2$.
In CPA the quantity $\Gamma$ ist determined self-consistently
in terms of the local Green's function
\mbox{$G_0(z)=\sum\limits_{k_xk_yk_z}[z+6-2(\cos k_x+\cos k_y+ \cos k_z)]^{-1}$}
of the ordered cubic lattice
(the sum runs over the Brillouin zone and
$z=-\omega^2+i\epsilon$) as follows:
\begin{equation}\label{e1}
\left\langle
\frac{
\Gamma(z)-K_{ij}
}{
1-
\left(
\Gamma(z)-K_{ij}
\right)
\left(
1-z\tilde G_0(z)
\right)/3\Gamma(z)
}
\right\rangle =0
\end{equation}
with $\tilde G_0(z)=G_0\mbox{{\large (}}z/\Gamma(z)
\mbox{{\large )}}/\Gamma(z)$
and \newline
\mbox{$\langle A \rangle =\int\mbox{d}K_{ij}P(K_{ij})A(K_{ij})$.}
The DOS is obtained from a numerical solution of eq. $\!$(\ref{e1})
and use of the relation
\mbox{$g(\omega )=-(2/\pi)\omega \mbox{Im}\{\tilde G(z)\}$.}

Preliminary calculations showed that without lower cutoff
(i. e. 
$K_{\mbox{{\footnotesize min}}}=-\infty $) the system becomes instable
for $\sigma \approx 0.6$. In order to be able to study a strongly disordered
system we have set $\sigma = 1$ and varied the cutoff
$K_{\mbox{{\footnotesize min}}}$
which controls the amount of small and negative force constants. 
Fig. 1 shows results of such calculations.
The excellent agreement of the CPA
calculations with the numerical analysis \cite{comment2}
(except for the immediate vicinity of the instability)
indicates both the reliability of the
CPA and the correctness of the procedure
utilized for the elimination of finite-size effects in the numerical work.
As $K_{\mbox{{\footnotesize min}}}$ decreases  from positive to negative
values, the peak in
$g(\omega )/\omega ^2$ shifts towards smaller frequencies.
At the same time the peak intensity increases.
In CPA
the system becomes instable for
$K_{\mbox{{\footnotesize min}}}=-0.85$, whereas in the numerical
calculation
the instability already occurs near
$K_{\mbox{{\footnotesize min}}}=-0.6$.
The low-frequency peak thus plays the role of a
precursor of the instability introduced by the presence of
small and negative force constants.
We therefore conclude that a low-frequency peak observed in
the reduced DOS of a disordered harmonic
system indicates that it is almost unstable.

Another quantity of interest, in particular in the context
of thermal conductivity data in
structural glasses \cite{and81,gga86}, is the mean free path of the phonons
$\ell $.
Within the CPA treatment
the mean free path can be identified with the decay length
of the wave intensity $|\exp\{i\omega r/v(\omega )\}|^2$ and is therefore
given by
\begin{equation}\label{e2}
\ell^{-1}(\omega)=
\frac{
2\omega \mbox{Im}\{v(\omega)\}
}{|v(\omega )|^2
}
\end{equation}

and behaves as $\ell^{-1}\propto \omega^4$ for $\omega \rightarrow 0$
(Rayleigh scattering).

In fig. 2 
we have plotted the reduced quantity
$\ell ^{-1}(\omega )/\omega^4$ for the same force constant distributions
as in fig. 1. It is seen that
the mean free path strongly decreases with the amount of small and
negative force constants.
In scattering experiments the decrease in $\ell (\omega )$
would show up as an increasing width of the
Brillouin peaks.
The $\omega^4$ dependence of $\ell(\omega)$ levels off
near the frequency where the
reduced DOS has its maximum.
It has already been demonstrated in the literature that such a
frequency dependence of $\ell $ is capable to explain the plateau
in the thermal conductivity \cite{and81,ws93}.
One sees from fig. 2 that for
$K_{\mbox{{\footnotesize min}}}=-0.6$ (almost unstable case)
$\ell $ is in this frequency range of order unity.
Since the real part of $v(\omega)$ and $\omega $ are of order unity
as well the mean free path becomes comparable to the wavelength,
in which case the phonons may become localized according
to the Ioffe-Regel criterion
\cite{gga86}.
 
Therefore we were interested in the question as to whether
in our model the vibrational states are localized. This can be
checked by an investigation of the level statistics
\cite{izr90}. If we define normalized eigenvalues
by $\epsilon_i=\bar F(\lambda_i)$ a histogram of the
distances $s_i=|\epsilon_i-\epsilon_{i+1}|$ should yield a
distribution according to the Gaussian orthogonal random matrix
ensemble (GOE) $P(s)=\frac{1}{2}\pi s\exp\{-\pi s^2/4\}$ in the
case of delocalized states, whereas one expects a Poissonian
distribution $P(s)=\exp\{-s\}$ for localized states. This is due
to the fact, that the delocalized states show level repulsion, whereas
the localized ones do not.

In fig. 3 we show a representative set of the level distance statistics
of our calculated spectra.
It is clearly seen that they universally follow
the GOE statistics. We therefore conclude that the states in our
model are all delocalized for the force constant distributions
considered, also in the vicinity of the instability.

When comparing our results with real three-dimensional disordered
systems like glasses a few comments are in order.
We have treated a scalar model and thus ignored the vector
character of the phonons completely. We believe, however,
that our model already is able to mimic
the behavior of transverse phonons
in a glass. If the vector character is taken into account, one
would expect an even stronger scattering due to the admixture
of longitudinal degrees of freedom. We therefore believe
that our model underestimates the intensity of the boson
peak somewhat. We expect also that in a modified version of our model
the states near the boson peak may become Anderson-localized near the
borderline of stability.
Therefore we have to state that in a real glassy system it still remains
an open question, whether the modes associated with the boson peak
are localized or delocalized. Our model shows, however, that is not
necessary to postulate the existence of localized states or
strongly anharmonic effects to obtain an excess contribution
in the DOS.

As we have treated only harmonic interactions the effects associated
with the glass-liquid transformation \cite{gs96,got91} are outside the
scope of the present model.

In conclusion we have solved a simple model of coupled harmonic
oscillators, both, numerically and in coherent-potential approximation.
Near the point of instability the model exhibits a low-frequency peak in
the reduced density of states $g(\omega )/\omega^2$,
which we view as the analogue to the boson peak observed in
structural glasses.

It is a pleasure to thank W. G\"otze, W. Petry, A. Meyer,
and J. Wuttke
for fruitful discussions and a critical reading of the manuscript.

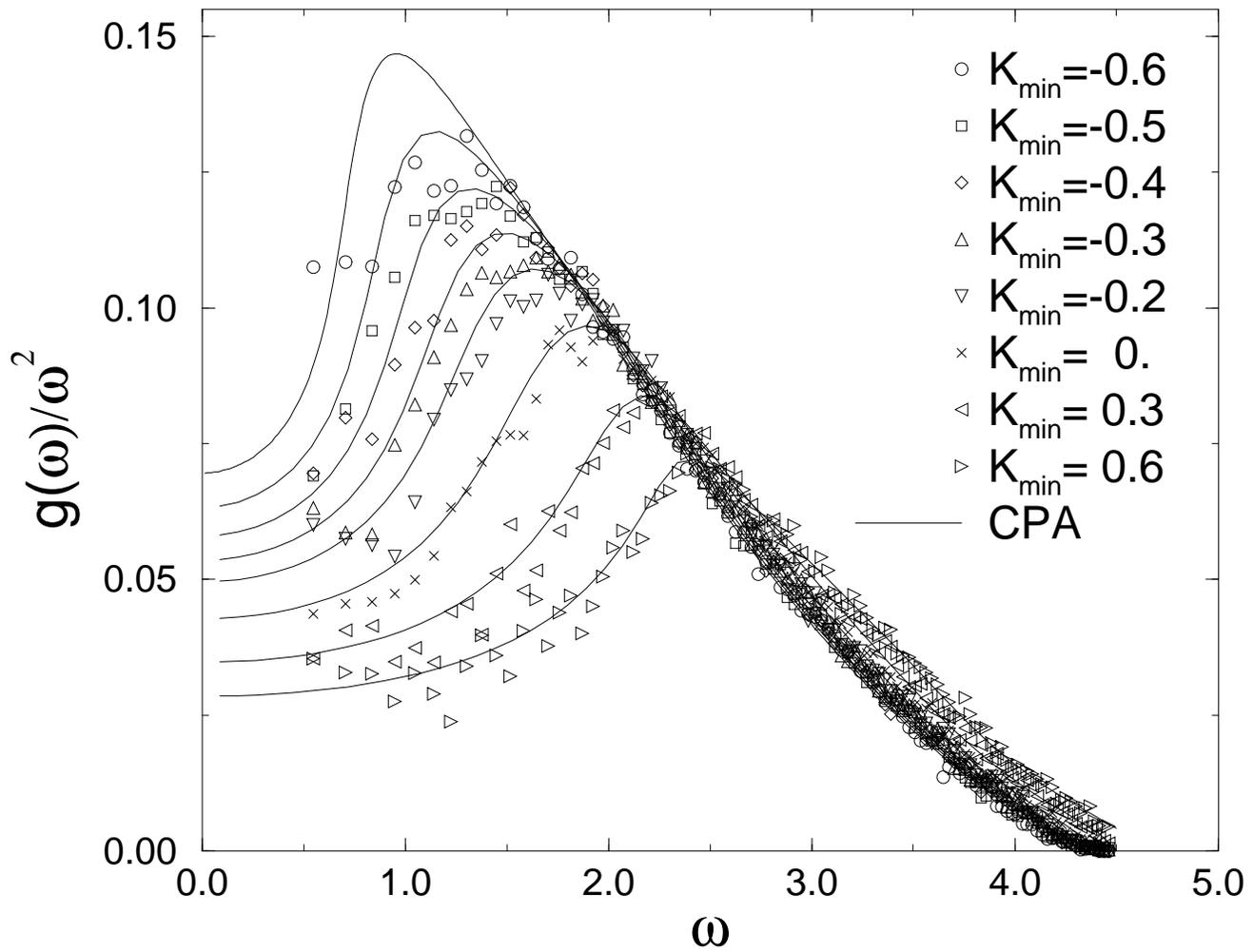
FIG. 1:\newline
Reduced DOS $g(\omega)/\omega^2$ against frequency for force constant 
distributions with $\sigma =K_0=1$ and different lower cutoffs
$K_{\mbox{{\footnotesize min}}}$. The symbols represent the numerical
diagonalization, the full lines the CPA results. The agreement is achieved
without any adjustable parameters.

\vspace{0.5cm}

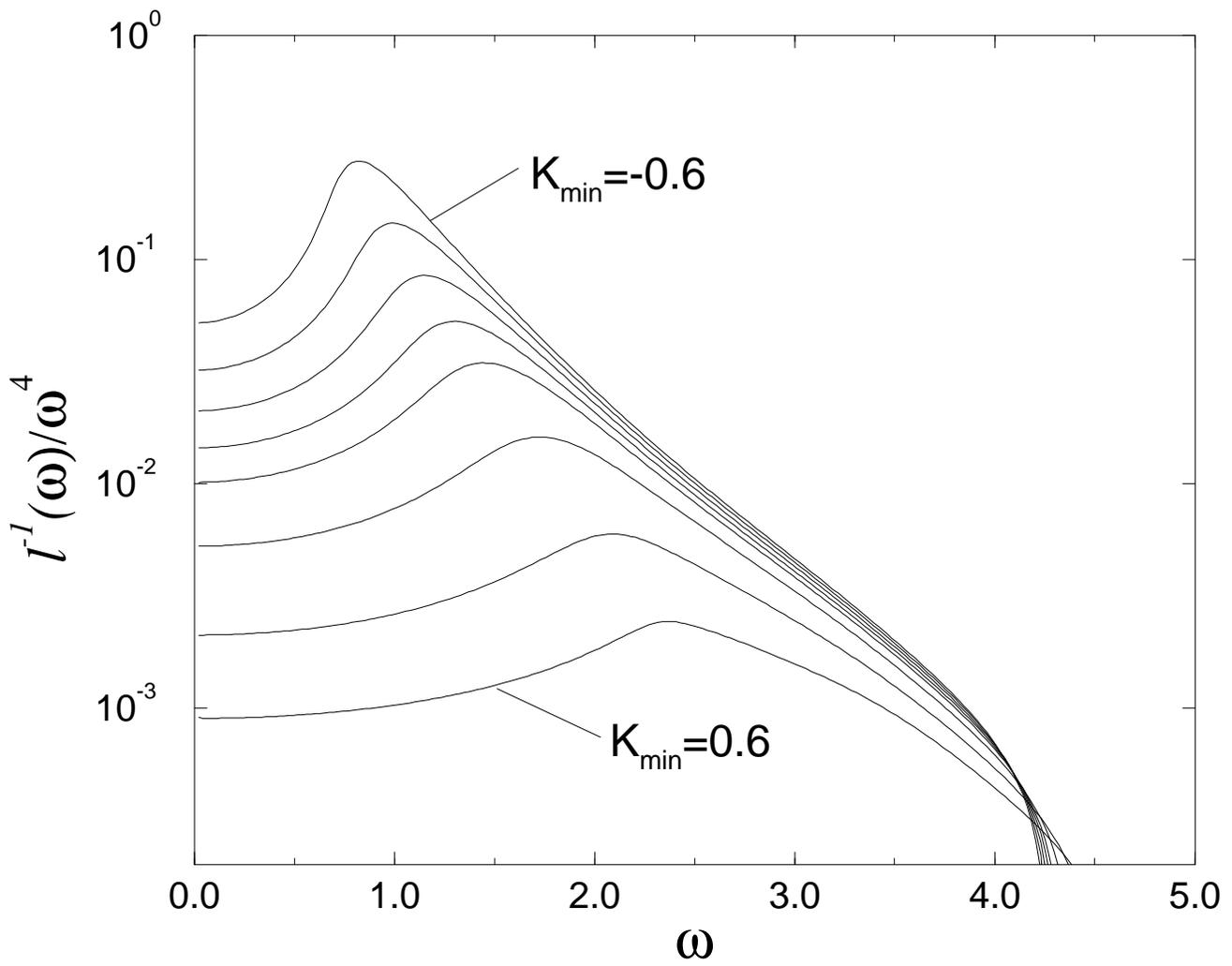
FIG. 2:\newline
Reduced inverse mean free path
$\ell^{-1}/\omega^4$
calculated in CPA according to (\ref{e2}) for the 
parameters of fig. 1. 

\vspace{0.5cm}

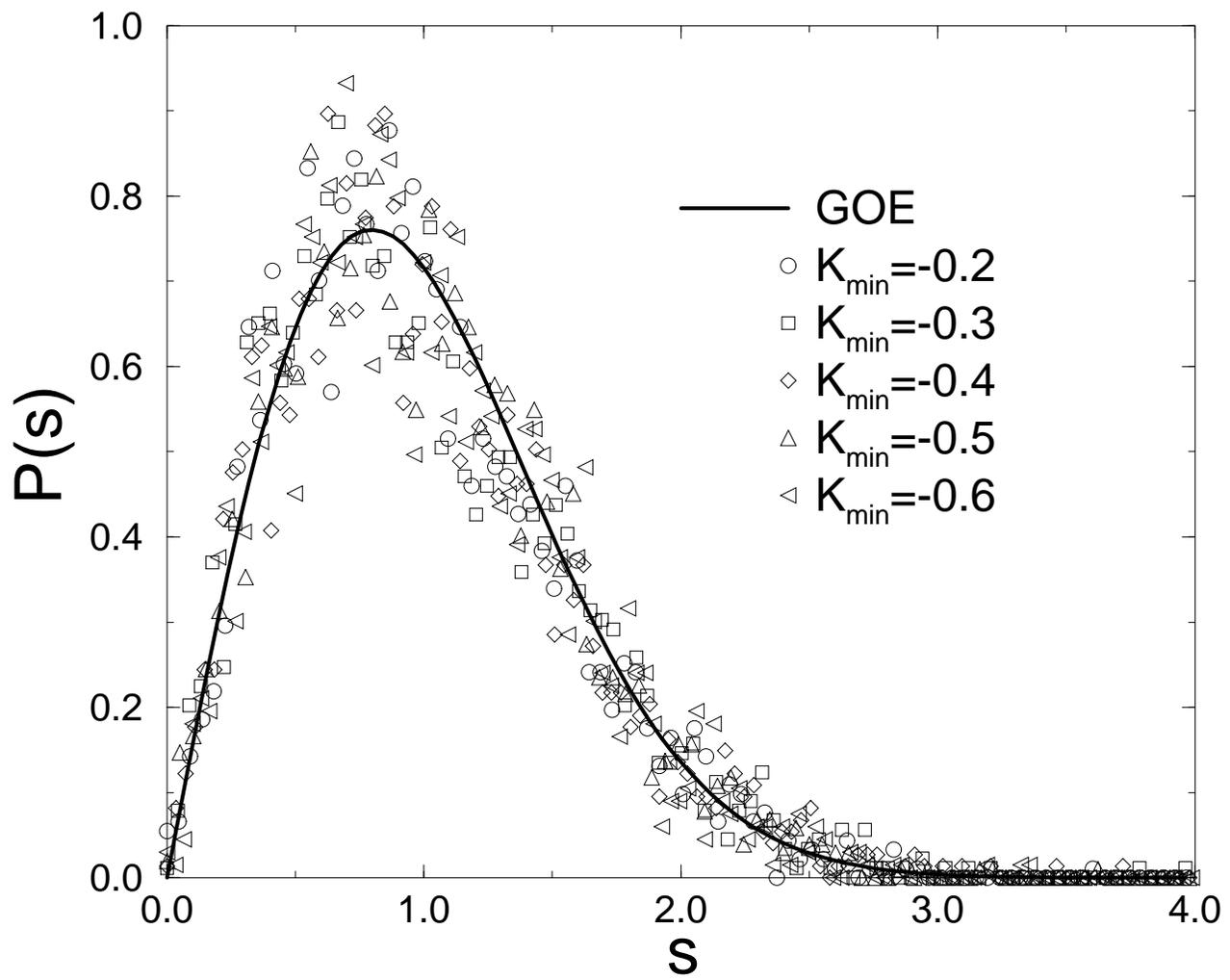
FIG. 3:\newline
Level distance statistics for the distributions with negative values of
$K_{\mbox{{\footnotesize min}}}$. Since they follow the statistics of the
Gaussian orthogonal ensemble (GOE), the corresponding states are delocalized.

\begin{references}
\bibitem{comment2} For the good agreement with the numerical calculation
it is essential to use the exact lattice Green's function $G_0(z)$
and not Hubbard's semicircular approximation 
which has been used in refs. \cite{doy84,ol81}.
\end{references}
\end{document}